%Paper: hep-th/9206046
%From: minahan@gomez.phys.virginia.edu
%Date: Wed, 10 Jun 1992 14:38:04 EDT

\input phyzzx
%\PHYSREV
%\normalspace
%
\Pubnum={UVA-HET-92-04\cr
CU-TP-566\cr
hepth@xxx/9206046}
\date={June 1992}
%\pubtype={Draft}
\titlepage
\title{Integrable Systems for Particles with Internal Degrees of Freedom}
\bigskip
\author {Joseph~A.~Minahan\footnote\dag
{minahan@gomez.phys.virginia.edu}}
\address{Department of Physics,  Jesse Beams Laboratory,\break
University of Virginia, Charlottesville, VA 22901 USA}
\andauthor{Alexios~P.~Polychronakos\footnote\ddag
{app@cuphyf.phys.columbia.edu}}
\address{Pupin Physics Laboratories, Columbia University,\break
New York, NY 10027}
\bigskip
\abstract{
We show that a class of models for particles with internal degrees of
freedom are integrable.
These systems are basically generalizations of the models of Calogero and
Sutherland.  The proofs of integrability are based on a
recently developed exchange operator formalism.
We calculate the wave-functions for the Calogero-like models and find the
ground-state wave-function for a Calogero-like model in
a position dependent magnetic field.  This last model might have some relevance
for matrix models of open strings.}
\submit{Physical Review Letters, {\rm PACS numbers: 03.65.Ca, 03.65.Ge}}
\endpage

\def\PL{{\it Phys. Lett.\ }}

\def\PRA{{\it Phys. Rev. A\ }}
\def\PRL{{\it Phys. Rev. Lett.\ }}

\REF\Calog{F.~Calogero, {\it Jour. of Math. Phys.}
{\bf10}, 2191 and 2197 (1969); {\bf 12}, 419 (1971).}
\REF\Moser{J.~Moser, {\it Adv. Math.} {\bf 16}, 1 (1975); F.~Calogero, {\it
Lett. Nuovo Cim.} {\bf13}, 411 (1975); F.~Calogero and C. Marchioro, {\it
Lett. Nuovo Cim.} {\bf13}, 383 (1975).}
\REF\Suth{B.~Sutherland, \PRA {\bf 4}, 2019 (1971); \PRA {\bf5}, 1372 (1972);
\PRL {\bf34}, 1083 (1975).}
\REF\HalShas{F.~D.~M.~Haldane, \PRL {\bf 60}, 635 (1988); B.~S.~Shastry,
\PRL {\bf 60}, 639 (1988).}
\REF\appext{A.~P.~Polychronakos, \PL {\bf B266}, 29 (1991); \PL {\bf B276},
341 (1992); \PL {\bf B277}, 102 (1992).}
\REF\HaHal{Z.~N.~C.~Ha and F.~D.~M.~Haldane, Princeton preprint, 1992.}
\REF\app{A.~P.~Polychronakos, Columbia preprint CU-TP-551, 1992, to appear
in Phys. Rev. Lett.}
\REF\jam{J.~A.~Minahan, University of Virginia preprint UVA-92-01, 1992, to
appear in Nucl. Phys. B.}

%A slew of definitions
\def\half{{1\over2}}
\def\al{\alpha}

\def\gam{\gamma}

\def\lam{\lambda}

\def\Jv{\vec J}

\def\Vij{V_{ij}}
\def\Mij{M_{ij}}
\def\pii{\pi_i}
\def\sigij{\sigma_{ij}}
\def\sigv{\vec \sigma}
\def\sigz{\sigma^z}
\def\Tij{T_{ij}}
\def\adag{a^\dagger}
\def\dij{\delta_{ij}}
\def\psigs{\psi_{\rm gs}}
\def\psiex{\psi_{\rm ex}}
\def\pit{\tilde\pi}
\def\It{\tilde I}
\def\Nf{N_f}

The integrable systems of Calogero, Moser and Sutherland are many body
systems in one dimension, in which $N$ identical
particles interact among themselves
with two-body inverse square potentials [\Calog-\Suth].
Other integrable systems have been discovered which are related to these
models.  These include the lattice version of Haldane and Shastry [\HalShas],
and more recently,
models where the particles are exposed to particular types of
external potentials [\appext].

In this paper we will further generalize
the Calogero-Moser-Sutherland models by introducing particles
with internal degrees of freedom.
We will generalize both the Calogero model
and the Sutherland model by including
terms in the hamiltonian that couple the internal degrees
of freedom.
We will then show that these generalized models are integrable.
We will also construct the wave-functions
of the ground state and excited states for
the Calogero-like model.  Wave-functions for the Sutherland-like model
were recently given [\HaHal] and won't be repeated here.
We will also consider a Calogero-like model where internal spins are
coupled to a magnetic field which is linear in position space.   We will
find the ground state wave function for this model,
although at this point, we don't know how to prove if the system is integrable.

In this paper we will make extensive use of the exchange operator
formalism developed by one of the authors [\app], which we review here.
Consider the generalized momentum operator $\pii$,
$$\pii=p_i+i\sum_{j\ne i}\Vij\Mij,\eqn\genmom$$
where $\Vij=V(x_i-x_j)$ and $\Mij$ is an exchange operator that interchanges
the positions of the $i$ and $j$ particles.  We will assume that it does not
exchange the internal quantum numbers of these particles.
In order for $\pii$ to be
self-adjoint, $V(x)$ must satisfy $V^\dagger(-x)=-V(x)$.

The simplest hamiltonian that can be constructed with the $\pii$ operators is
$$H=\half\sum_i\pii^2,\eqn\freeham$$
which can be expanded to
$$H=\half\sum_i p_i^2+\half\sum_{i\ne j}\biggl[\Vij(p_i+p_j)\Mij+V'_{ij}\Mij
+\Vij^2\biggr]-{1\over6}\sum_{i\ne j\ne k\ne i}V_{ijk}M_{ijk}.\eqn\hamexp$$
$V_{ijk}$ is given by
$$V_{ijk}=\Vij V_{jk}+V_{jk}V_{ki}+V_{ki}\Vij,\eqn\vijkeq$$
and $M_{ijk}$ is the three body exchange operator,
$$M_{ijk}=M_{ij}M_{jk}=M_{jk}M_{ki}=M_{ki}M_{ij}.\eqn\Mijkeq$$
If $V(x)=-V(-x)$ then the terms linear in $p_i$ drop out of the hamiltonian
and the commutation relations for $\pii$ satisfy
$$[\pii,\pi_j]=\sum_{k\ne i,j}V_{ijk}(M_{ijk}-M_{jik}).\eqn\picomm$$

Let us now assume that $V(x)=l/x$, in which case $V_{ijk}=0$.  Therefore
$\pii$ commutes with $\pi_j$ and hence with $H$.  Furthermore, we can
construct totally symmetric operators
$$I_n=\sum_i\pii^n\eqn\Indef$$
which commute with $H$ as well as with themselves.  Since we can
construct $N$ local conserved quantities, the system is integrable.

Let us now include the internal degrees of freedom.
Assume that the particles are labeled by an index that transforms
under the fundamental representation of
$SU(\Nf)$.   Let $\sigij$ be the operator
that exchanges the indices of particles $i$ and $j$.
The $\Nf\times \Nf$ matrices $\tau^a$, which are the
generators of $SU(\Nf)$ in the fundamental representation,
satisfy the identity
$$\sum_a\tau^a_{\mu\nu}\tau^a_{\lam\rho}=\half
\Bigl(\delta_{\mu\rho}\delta_{\nu\lam}
-{1\over \Nf}\delta_{\mu\nu}\delta_{\lam\rho}\Bigr).\eqn\taueq$$
It is then straightforward to check that
$$\sigij={1\over \Nf}+2\sum_a\tau^a_i\otimes\tau^a_j,\eqn\spinx$$
where $\tau^a_i$ acts on the $i^{\rm th}$ particle.
We could also consider higher spin representations, but then the exchange
operator is a nonlinear expression of the $SU(N)$ generators.

We also define the total exchange operator, $\Tij=\Mij\sigij$, which
exchanges both positions and $SU(\Nf)$ indices.
For a system of $N$ identical particles, the wave functions will be
eigenfunctions of $\Tij$.  Since $\pii$ has no dependence on the $SU(\Nf)$
indices,
$$[\pii,\sigma_{jk}]=0.$$
Thus, the integrability is preserved, but the introduction of internal
indices will modify the effect of the $\Mij$ on the wave functions.
Keeping $V(x)=l/x$, the hamiltonian \hamexp\ reduces to
$$H=\half\sum_i p_i^2+\half\sum_{i\ne j}{l(l-\Mij)\over(x_i-x_j)^2}.
\eqn\hamred$$
Since $\Mij^2=\sigij^2=1$, we can write $\Mij=\Tij\sigij$.  If we assume that
$H$ is acting on a system of identical fermions, then $H$ can be rewritten as
$$H=\half\sum_i p_i^2+\half\sum_{i\ne j}{l(l+1/\Nf+
2\sum_a\tau^a_i\otimes\tau^a_j)
\over(x_i-x_j)^2}. \eqn\hamspin$$
If the internal group is $SU(2)$ and $l=-1$,
then the interaction term is precisely the term that has
arisen recently in the study of matrix models for one-dimensional
open strings [\jam].

The wave functions for the hamiltonian in \hamspin\ are not normalizable,
so instead
let us consider the $SU(\Nf)$ generalization of Calogero's model of interacting
particles which reside in a central harmonic potential.  To this end,
consider the operator
$$h_i=\adag_i a_i=(\pi_i+i\omega x_i)(\pii-i\omega x_i).\eqn\heqn$$
The commutator of $x_i$ with $\pi_j$ is
$$[x_i,\pi_j]=i\dij(1+l\sum_{k\ne i} M_{ik})-i(1-\dij)\Mij.\eqn\xpicomm$$
Using this relation, one can derive the following:
$$\eqalign{[a_i,a_j]&=[\adag_i,\adag_j]=0\cr
[a_i,\adag_j]&=2\omega\dij(1+l\sum_{k\ne i} M_{ik})-2l\omega(1-\dij)\Mij\cr
[h_i,h_j]&=-2l\omega(h_i\Mij-\Mij h_i)\cr
[h_i^n,h_j]&=-2l\omega(h_i^n\Mij-\Mij h_i^n).}\eqn\commrel$$
Let the hamiltonian be given by
$$\eqalign{H&=\half\sum_i h_i+{l\omega\over2}\sum_{i\ne j} \Mij\cr
&= \half\sum_i (p_i^2+\omega^2x_i^2)
+\half\sum_{i\ne j}{l(l-\Mij)
\over(x_i-x_j)^2}+{N\omega\over2}.}\eqn\harmham$$
This hamiltonian is slightly different than the one in ref. [\app] and is used
to accomodate the fact that the wavefunctions are no longer eigenvalues
of $\Mij$.
Using the commutation relations in \commrel\ we find that
$$\eqalign{[H,a_j]&=-\omega a_j\cr
[H,\adag_j]&=\omega\adag_j.}\eqn\hcomma$$

We briefly repeat the proof in ref. [\app] that shows this system is
integrable.  Consider the sums
$$I_n=\sum_ih_i^n.\eqn\Ineq$$
Computing the commutators of the $I_n$, we find
$$\eqalign{[I_n,I_m]&=\sum_{i,j}[h_i^n,h_j^m]=\sum_{i,j}\sum_{\al=0}^{m-1}
h_j^\al[h_i^n,h_j]h_j^{m-\al-1}\cr
&=-2l\omega\sum_{i,j}\Biggl(\sum_{\al=0}^{m-1}-\sum_{\al=n}^{m+n-1}\Biggr)
h_j^\al\Mij h_j^{m+n-\al-1}.}\eqn\Incomm$$
Explicitly antisymmetrizing in $m$ and $n$ then gives
$$[I_n,I_m]=-l\omega\sum_{i,j}\Biggl(\sum_{\al=0}^{m-1}-\sum_{\al=n}^{m+n-1}
-\sum_{\al=0}^{n-1}+\sum_{\al=m}^{m+n-1}\Biggr)h_j^\al\Mij h_j^{m+n-\al-1}=0.
\eqn\Incommz$$
All $I_n$ commute with the hamiltonian, hence the system is integrable.

Now let us bring the internal indices into the picture.
For what follows we will assume that the particles transform
under the spin $1/2$ representation of $SU(2)$.
Since $\sigv_i$ commutes
with all $I_n$, the system remains integrable, but again
the wave functions are, in general,
no longer eigenvalues of $\Mij$.  Because of the commutation
relations in \hcomma, in order that the spectrum be bounded from below,
the ground state should be annihilated by all $a_i$.  Since $a_i$ has no
spin dependence, the ground-state wave function can be expressed as a product
of spatial and spinor wave functions.
For $N$ identical fermions
the wave function is totally symmetric under spin-exchange,
hence the total spin of the system transforms under a spin $N/2$
representation.  Therefore, the ground state is $N$-fold degenerate and its
wave functions are given by
$$\psi_{\rm gs}^m=\prod_{i<j}(x_i-x_j)|x_i-x_j|^{-1-l}e^{-\half\sum_i\omega
x_i^2}\chi_m(\sum_j\sigma_j),\eqn\psigseq$$
where $m$ is the total spin along the $z$ direction and
$$\eqalign{\chi_m(J)&=1\qquad {\rm if}\qquad J=m,\cr
&=0\qquad {\rm otherwise.}}\eqn\chimeq$$

All excited states can be constructed by acting on the ground
state with symmetric combination of $\adag_i$ operators.
For instance, if we act on the gound state with $\sum(\adag_i)^n$, we create
a state with energy $n\omega$ above the ground state.  Since $\sigij$
commutes with this operator, the total spin of this state is still in an
$N/2$ representation.
The only other set of operators that we need to create excited states are
of the form
$\sum_i\sigv_i(\adag_i)^n$.
(If we were considering a generic internal group, we could substitute
any element of the Lie algebra for $\sigv_i$.)
This operator does not commute with $(\sum\sigv_i)^2$, so this will
create excited states in representations different from $N/2$.

Acting on the ground state with $\sum\sigv_i(\adag_i)^n$ results in
the wave function
$$\eqalign{\psiex^m&=\sum_i\sigv_i(\adag_i)^n\psigs\cr
&=\biggl[{1\over N}\Bigl(\sum_j(\adag_i)^n\Bigr)\Bigl(\sum_i\sigv_i\Bigr)
\psigs^m\biggr]+
\biggl[{1\over 2N}\sum_{i,j}((\adag_i)^n-(\adag_j)^n)(\sigv_i-\sigv_j)
\psigs^m\biggr]\cr
&=\psi_1^{m}+\psi_2^{m}.}
\eqn\psiexeq$$
The first term in square brackets ($\psi_1^{m}$) is
clearly a component of a spin $N/2$ representation.
On the other hand, the term in the second set of square brackets
($\psi_2^{m}$) is
a spin $N/2-1$ component.  One can verify this by noting that
$$(\Jv)^2=(\sum_i\sigv_i)^2=\sum_{i<j}\sigij-{N(N-4)\over4}.\eqn\tspin$$
It is then straightforward to show that
$$\sum_{k<l}\sigma_{kl}((\adag_i)^n-(\adag_j)^n)(\sigv_i-\sigv_j)\psigs^m=
{N(N-3)\over2}((\adag_i)^n-(\adag_j)^n)(\sigv_i-\sigv_j)\psigs^m.\eqn\updown$$
Therefore,
$$(\Jv)^2\psi_2^{m}
={N\over2}\big({N\over2}-1\big)\psi_2^{m}.\eqn\tspineq$$
and hence, $\psi_2^{m}$ is the wave function for a spin $N/2-1$ state.
We can act on the ground state with $M$ copies of $\sum\sigv_i(\adag_i)^n$
which create states at energies $Mn\omega$ above the ground state.  The
resulting wave function will be a sum of wave functions with definite total
spin, with the lowest total spin being $(N/2-M)$, if $M\le N/2$.  In fact,
the lowest energy for a state in an $(N/2-M)$ representation is
$$E_{0,M}=M\omega +{N\omega\over2}-{N(N-1)l\omega\over2}.\eqn\EMlow$$

Finally, let us consider the generalized Sutherland model, which has recently
been discussed in ref. [\HaHal].  In this case, we choose
$V(x)=l\cot ax$, which gives $V_{ijk}=l^2$ and leads to the
commutation relations
$$[\pi_i,\pi_j]=l^2\sum_k(M_{ijk}-M_{jik}).\eqn\Suthcomm$$
We define the hamiltonian to be
$$H=\half\sum_i\pi_i^2+{1\over6}\sum_{i\ne j\ne k\ne i}M_{ijk}
=\half\sum_ip_i^2+\half\sum_{i\ne j}{l(l-\Mij)\over\sin^2(x_i-x_j)}.
\eqn\Suthham$$
The commutator of $H$ with $\pi_j$ is given by
$$\eqalign{[H,\pi_j]&=[\half\sum_i\pi_i^2+{1\over6}\sum_{i\ne m\ne k\ne i}
M_{imk},\pi_j]\cr
&=l^2\sum_{i\ne k}\Bigl(\pi_i(M_{ijk}-M_{jik})+(M_{ijk}-M_{jik})\pi_i\Bigr)\cr
&\qquad\qquad+l^2\sum_{i\ne k}(M_{ijk}\pi_j-\pi_jM_{ijk})\cr
&=l^2\sum_{i\ne k}(\pi_i+\pi_k)(M_{ijk}-M_{jik})=0,}\eqn\Hampi$$
where the final sum is zero because of the explicit antisymmetry between
$i$ and $k$.  To complete the proof of integrability [\app],
consider the modified momentum operator $\pit_i$,
$$\pit_i=\pi_i +l\sum_{j\ne i}\Mij,\eqn\piteq$$
which also commutes with $H$.  $\pit_i$ satisfies the commutation relations
$$[\pit_i,\pit_j]=2l(\pit_i\Mij-\Mij\pit_i)\eqn\pitcomm$$
which are of the same form as the commutation relations for the $h_i$ operators
in the Calogero case.  Therefore, the operators $\It_n=\sum\pit_i^n$ form
a commuting set that also commute with the Hamiltonian and the system is
integrable.

All of these models remain integrable if a spatially constant external field
coupled to the internal indices
is included in the hamiltonian.  However, the hamiltonian derived
for the open string matrix model [\jam] contains a position dependent
magnetic field which couples to the spins.
One consequence of this is that the total spin no longer
commutes with the Hamiltonian.
While we presently don't know how to show if such models are integrable, we
have succeeded in finding the ground state wave function for one such model.
Its hamiltonian is given by
$$\eqalign{H&= \half\sum_i (p_i^2+\omega^2x_i^2+2\gam x_i\sigz_i)
+\half\sum_{i\ne j}\biggl({l(l-\Mij)\over(x_i-x_j)^2}+{l\omega}
\sigij\biggr)
-N({\omega\over2}-{\gam^2\over8\omega^2})\cr
&=\half\sum_i\adag_ia_i+{l\omega\over2}\sum_{i\ne j}(\Mij+\sigij),}
\eqn\harmhammag$$
where $a_i$ is defined as
$$a_i=\pi_i-i(\omega x_i+{\gam\over\omega}\sigz_i).\eqn\aidef$$
A short calculation shows that
$$\eqalign{[H,a_i]&=-\omega a_i
+il \gam\sum_{j\ne i}(\sigz_i-\sigz_j)(\Mij+\sigij),\cr
[H,\adag_i]&=+\omega \adag_i
+il \gam\sum_{j\ne i}(\sigz_i-\sigz_j)(\Mij-\sigij).}
\eqn\Hcommai$$
While the commutators in \Hcommai\ don't have the usual form for
creation and annihilation operators, one can observe that if the hamiltonian
in \harmhammag\ acts on an antisymmetric state, then it reduces to
a positive definite operator.  Therefore, any antisymmetric state annihilated
by all $a_i$ must be a ground state.
The ground state has an $N$-fold degeneracy, with the states labeled by
the total spin in the $z$ direction, and it is straightforward to show
that the wave functions are given by
$$\psigs^m=\sum_{i<j}(x_i-x_j)|x_i-x_j|^{-1-l}e^{-\half\sum_i\omega(x_i+\gam
\sigz_i/\omega)^2}\chi_m(\sum_j\sigz_j).\eqn\gseqnew$$
Proof of integrability requires further work.

\ack{
J.A.M. was supported in part by
D.O.E. grant DE-AS05-85ER-40518.  A.P.P. was supported
in part by D.O.E. grant DE-AC02-76ER-02271.}

\refout
\end